%% file: main.tex
\documentclass[prd, 11pt,nofootinbib, superscriptaddress, preprintnumbers,floatfix]{revtex4}
\usepackage[utf8]{inputenc}
\usepackage{amsmath,amssymb}
\usepackage{graphicx}
\usepackage{subfigure}
\usepackage{hyperref}
\usepackage{nicefrac}
\usepackage{color}
\usepackage{dsfont}
\usepackage{placeins}

\DeclareMathOperator*{\tr}{Tr}
\DeclareMathOperator*{\re}{Re}
\DeclareMathOperator*{\im}{Im}

\newcommand{\sd}{\operatorname{sd}}
\newcommand{\fig}[1]{Figure~\ref{#1}}
\newcommand{\tab}[1]{Table~\ref{#1}}
\newcommand{\eq}[1]{Eq.~\ref{#1}}
\newcommand{\eqs}[1]{Eqs.~\ref{#1}}
\newcommand{\h}{\mathcal{H}}
\newcommand{\s}{\mathcal{S}}
\newcommand{\dH}{\Delta\mathcal{H}}
\newcommand{\ddH}{\delta\Delta\mathcal{H}}
\newcommand{\su}{\mathrm{SU}(2)}
\newcommand{\ii}{\mathrm{i}}
\newcommand{\e}{\mathrm{e}}

\graphicspath{{plots/}}

\begin{document}

\title{Reversibility Violation in the Hybrid Monte Carlo Algorithm}

\newcommand\bn{HISKP (Theory) and BCTP, University of Bonn, Bonn, Germany}
\newcommand\rom{Dipartimento di Fisica, Universit{\`a} and INFN di Roma Tor Vergata, 00133 Roma, Italy}
\newcommand\bern{Albert Einstein Center for Fundamental Physics, University of Bern, 3012 Bern, Switzerland}
\newcommand\cypa{Department of Physics, University of Cyprus, PO Box 20537, 1678 Nicosia, Cyprus}
\newcommand\cypb{Computation-based Science and Technology Research Center, The Cyprus Institute}
\newcommand\wup{Fakult\"at f\"ur Mathematik und Naturwissenschaften, Bergische Universit\"at Wuppertal}
\newcommand\roma{Centro Fermi - Museo Storico della Fisica e Centro Studi e Ricerche
Enrico Fermi, Compendio del Viminale, Piazza del
Viminiale 1, I-00184, Rome, Italy}
\newcommand\romb{Dipartimento di Fisica, Universit{\`a} di Roma ``Tor Vergata",
Via della Ricerca Scientifica 1, I-00133 Rome, Italy}

\author{Carsten Urbach}\affiliation{\bn}


\begin{abstract}
We investigate reversibility violations in the Hybrid Monte Carlo
algorithm. Those violations are inevitable when computers with finite
numerical precision are being used. In SU$(2)$ gauge theory, we study
the dependence of observables on the size of the reversibility
violations. While we cannot find any statistically significant
deviation in observables related to the simulated physical model,
algorithmic specific observables signal an upper bound for
reversibility violations below which simulations appear
unproblematic. This empirically derived condition is independent of
problem size and parameter values, at least in the range of parameters
studied here.
\end{abstract}

\maketitle


\input{intro}

\input{hmc}

\input{toy}

\input{results}

\input{summary}

\begin{acknowledgments}

The author
thanks K.~Jansen for injecting the idea for this project a long time
ago, for stimulating discussions and all his support. Thanks also to
B.~Kostrzewa and M.~Ueding for discussions and cross-checks and to
U.-G.~Meißner for useful comments on the draft.
The computer time for this project was made available to us in parts
by the John von Neumann-Institute for Computing (NIC) on the Jureca
system in J{\"u}lich. 
This project was funded by the DFG as a project in
the Sino-German CRC110. 
The open source software
package
R~\cite{R:2005} has been used.
\end{acknowledgments}

\input{appendix}

\FloatBarrier
\bibliographystyle{h-physrev5}
\bibliography{bibliography}

\end{document}

%% file: intro.tex
\section{Introduction}

The Hybrid Monte Carlo (HMC) algorithm~\cite{Duane:1987de} is an exact
accept/reject Markov chain Monte Carlo algorithm. It allows one to
perform global updates combined with large acceptance rates. This
property makes the HMC in its
variants~\cite{Frezzotti:1997ym,Clark:2004cp,Montvay:2005tj} and with its
improvements~\cite{Hasenbusch:2001ne,Hasenbusch:2002ai,Luscher:2005rx,Urbach:2005ji} the 
workhorse for lattice 
Quantum Chromodynamics (QCD) simulations with dynamical fermions. 

The HMC is composed of a molecular dynamics (MD) update and a Metropolis
accept/reject step. During the MD update, Hamilton's equations of motion
(EOM) are integrated, in practice numerically. The accept/reject step
corrects for finite integration step errors and renders the HMC
exact. However, the proof of exactness requires the numerical
integration scheme to be reversible and integration measure
conserving. Numerical integration schemes conserving the integration
measure are so-called symplectic 
integration schemes. A sub-set of these is also reversible, with the
\emph{leapfrog} integration scheme as the most well-known example. 

Any practical realisation of such integration schemes suffers
from round-off errors due to finite precision available on
computers. In fact, it has been known since a long time that reversibility
is violated in HMC simulations of lattice
QCD~\cite{Edwards:1996vs,Jansen:1995gz,Liu:1997fs}. Even further, the
underlying equations of motion are chaotic in nature. Thus, any small
round-off error will magnify exponentially during the integration. A
corresponding positive Lyapunov exponent can be determined.
Even though one may argue that these reversibility violations are a
property of the algorithm, and not of the simulated system, 
it was conjectured~\cite{Edwards:1996vs} that this Lyapunov
exponent obeys a continuum limit approached in a certain functional
form with the coupling constant of the theory.

However, this hypothesis has never been finally verified or
falsified. And, more importantly for practical simulations, to the
knowledge of the author it has never been checked whether or not
reversibility violations have any impact on observables. Analytic 
predictions are difficult here, because from a principle point of view
the proof of exactness is no longer applicable once reversibility
violations are present.

In this paper we are going to present an investigation of this issue
in SU$(2)$ gauge theory as a model. SU$(2)$ gauge theory shares many
properties with QCD, most importantly asymptotic freedom and
confinement, but it requires much less computer resources than
SU$(3)$, not to speak about the inclusion of dynamical
fermions. Therefore, we are able to study volume and lattice spacing
dependencies.

This allows us to derive an empirical condition for how large
reversibility violations appear tolerable in SU$(2)$ gauge theory. It
remains to be seen how this condition applies in case of QCD with
SU$(3)$ gauge fields and dynamical fermions.

In this paper we first describe the HMC algorithm followed by a
description of SU$(2)$ lattice gauge theory. Next we present results
and finish with a discussion and summary. Most of the data tables can
be found in the appendix.

%% file: hmc.tex
\section{The Hybrid Monte Carlo Algorithm}

Assume we are after sampling field variables $\phi=\{\phi_x\}$, with
$x$ being a multi-index not further specified at this level, from a
distribution
\begin{equation}
  \phi\quad\sim\quad e^{-\s(\phi)}\,.
\end{equation}
We call $\s\in\mathbb{R}$ the action, which is bounded from below. For the HMC
one introduces auxiliary variables $\pi=\{\pi_x\}$ as conjugate
momenta to the field variables $\phi$ and an artificial Hamiltonian
\begin{equation}
  \h[\pi, \phi]\ =\ \frac{1}{2}\pi^2 + S(\phi)\,.
\end{equation}
$\h$ is conserved under Hamilton's equation of motion (EOM).
Defining $z=(\pi,\,\phi)$, these EOMs may be written in the form
\begin{equation}
  \label{eq:eom}
  \dot z\ =\ \mathbb{J}\cdot\frac{\partial \h[z]}{\partial z}\,,\qquad
  \mathbb{J}\ =\ 
  \begin{pmatrix}
    0 & -\mathds{1} \\
    \mathds{1} & 0 \\
  \end{pmatrix}\,,
\end{equation}
with $\mathds{1}$ being unit matrices with dimension of $x$. 
In this form the symplectic structure of the EOMs becomes apparent. 
The dot notation represents time derivatives in an                                                
artificial HMC time $\tau$.
The HMC evolution starting from $\phi$ to $\phi^\prime$ is then
defined as follows:
\begin{enumerate}
\item Generate momenta $\pi$ from a standard normal distribution.
\item Evolve $z\equiv z(0)$ in HMC time using \eq{eq:eom} for a 
  trajectory of length $\tau$ to arrive at $z(\tau)$. We denote this
  time evolution with $\mathcal{T}_\mathcal{I}(\tau)$ for integrator
  $\mathcal{I}$, such that 
  \begin{equation}
    z(\tau)\ =\ \mathcal{T}_\mathcal{I}(\tau)\, z(0)\,.
  \end{equation}
\item Compute
  \begin{equation}
    \dH\ =\ \h[z(\tau)] - \h[z(0)]\,.
  \end{equation}
\item Accept $z(\tau)$ with probability
  \begin{equation}
    P_\mathrm{acc}\ =\ \min\{1,\ \exp(-\dH)\}\,.
  \end{equation}
  If accepted set $\phi^\prime = \phi(\tau)$, else $\phi^\prime = \phi(0)$.
\item restart at 1. with $\phi=\phi^\prime$. 
\end{enumerate}
Reversibility of an integration scheme $\mathcal{I}$ can now be
written as
\begin{equation}
  \label{eq:reversiblity}
  \mathcal{T}_\mathcal{I}(-\tau)\, \mathcal{T}_\mathcal{I}(\tau)\, z(0)\
  =\  \mathcal{T}_\mathcal{I}(\tau)\, \mathcal{T}_\mathcal{I}(-\tau)\,
  z(0)\ =\ z(0)\,.
\end{equation}
For the integration measure to be conserved the Jacobi determinant of
$\mathcal{T}_\mathcal{I}$ must be one. This is always the case if
$\mathcal{T}_\mathcal{I}$ is symplectic. For an elementary and nicely
accessible proof see Ref.~\cite{Rim:2017}.

In practice, the integration is performed with finite precision
$\epsilon$. Hence,
\begin{equation}
  \mathcal{T}^\epsilon_\mathcal{I}(-\tau)\, \mathcal{T}^\epsilon_\mathcal{I}(\tau)\, z(0)
  = z(0) + \delta z(\epsilon)\,.
\end{equation}
In order to measure reversibility violations in an actual simulation
one defines
\begin{equation}
  \ddH\ =\ \h\left[\mathcal{T}^\epsilon_\mathcal{I}(-\tau)\,
  \mathcal{T}^\epsilon_\mathcal{I}(\tau)\, z\right] - \h[z]\,.
\end{equation}
%
A well known and very useful property of the HMC algorithm is
\begin{equation}
  \label{eq:edH}
  \langle \exp(-\dH)\rangle \ =\ 1\,,
\end{equation}
which follows analytically from the measure being
conserved. Using this and the convexity of the exponential function it
follows 
\begin{equation}
  \label{eq:edH2}
  \exp(-\langle \dH\rangle)\ \leq\ 1\quad\Rightarrow\quad
  \langle\dH\rangle \geq 0\,.
\end{equation}
Here, $\langle .\rangle$ denotes the ensemble average over all
generated $z$. 
We note in passing that symplecticity of the integration scheme
implies the existence of a so-called shadow Hamiltonian which is
exactly conserved under time evolution $\mathcal{T}$ (see
e.g. Ref.~\cite{Kennedy:2012gk}).

Reversible integration schemes can be constructed to any order $n$ in
the discretisation error $\delta\tau^n$. The leapfrog (LF) is a second
order integration scheme reading
\begin{equation}
  \mathcal{T}_\mathrm{LF}(\delta\tau)\,z(0)\ =\ 
  \begin{cases}
    \phi(\delta\tau)\ =\ \phi(0) + \delta\tau\,\pi(\delta\tau/2)\\
    \pi(\delta\tau)\ =\ \pi(\delta\tau/2)- \frac{\delta\tau}{2}
    \frac{\partial\s[\phi(\delta\tau)]}{\partial\phi(\delta\tau)}\\
  \end{cases}\,,
\end{equation}
with
\begin{equation*}
  \pi(\delta\tau/2)\ =\ \pi(0) - \frac{\delta\tau}{2}
  \frac{\partial\s[\phi(0)]}{\partial\phi(0)}\,.
\end{equation*}
It represents a semi-implicit integration scheme and is symmetric around
$\delta\tau/2$. In addition 
to the LF integration scheme we will use a fourth order integration
scheme which we will conventionally denote as OMF4. Its details can be
found in Ref.~\cite{Omelyan:2003}.

%% file: toy.tex
\section{The Toy Model: SU$(2)$ Gauge Theory}

We are going to work on a discrete and finite space-time lattice 
\begin{equation}
  V^\Lambda\ =\ (L_s/a)^3 \times L_t/a\ \equiv\ L^3 \times T
\end{equation}
with a lattice spacing denoted as $a$ and periodic boundary
conditions. Hence, the possible set of coordinates is given as
\begin{equation}
  \Lambda\ =\ \{x = (x_0, x_1, x_2, x_3) :\ x_0=0,\ldots T-1,
  x_{1,2,3} = 0, \ldots L-1 \}\,.
\end{equation}
We introduce so-called link variables $U_\mu(x)\in\mathrm{SU}(2)$
connecting points $x$ and $x+a\hat\mu$, where $\hat\mu$ is the unit
vector in direction $\mu\in 0,1,2,3$. For the discretised action we
are going to use the Wilson plaquette gauge action reading
\begin{equation}
  \label{eq:actionsu2}
  \mathcal{S}[U]\ =\ \frac{\beta}{2}\, a^4\sum_{x\in\Lambda}\sum_{\mu<\nu}
  \re\,\tr\left[\mathds{1}_{2}-U_{\mu\nu}(x)\right]
\end{equation}
with plaquette variables
\begin{equation}
  \label{eq:plaquette}
  U_{\mu\nu}(x)\ =\ U_\mu(x)\, U_\nu(x+a\hat\mu)\,
  U^\dagger_\mu(x+a\hat\nu)\, U^\dagger_\nu(x)\,.
\end{equation}
$\beta=4/g_0^2$ is the inverse squared gauge coupling and $g_0$ the gauge
coupling. 

For the actual implementation it is used that any $U\in\mathrm{SU}(2)$
can be written as 
\begin{equation}
  \label{eq:su2rep}
  U\ =\
  \begin{pmatrix}
    a & b \\
    -b^\star & a^\star\\
  \end{pmatrix}\qquad\textrm{with}\quad aa^\star +
  bb^\star\ =\ 1\,,\quad a,b\in\mathbb{C}\,,
\end{equation}
which is a consequence of SU$(2)$ being homeomorphic to $S^3$. Using
Pauli matrices $\vec\sigma$, we may also write
\begin{equation}
  \label{eq:su2pauli}
  U\ =\ x_0\mathds{1}_2\ + \ii\, \vec x\, \vec \sigma
\end{equation}
with $(x_0, \vec x)\in S^3$. This allows one to identify
\begin{equation}
    x_0 = \re (a)\,,\ x_1 = \im (b)\,,\ x_2 = \re (b)\,,\ x_3 =
    \im(a)\,.
\end{equation}
The trace of an $\su$ matrix is directly given by
\begin{equation}
  \tr\, U\ =\ \tr\, U^\dagger\ =\ 2\,\re(a)\,.
\end{equation}
The representation \eq{eq:su2pauli} 
is efficiently used in a numerical implementation, since only four
real numbers need to be stored. One could reduce to only three real 
numbers, if $\det(U)=1$ was used as well.

Using the Pauli matrices we can now introduce the derivative of a
function $f(U), U\in\mathrm{SU}(2)$ as follows
\begin{equation}
  D_{j}f(U)\ =\ \frac{\partial}{\partial\alpha}
  f(\e^{\ii\alpha\sigma_j}U)|_{\alpha=0}\,,\qquad j=1,2,3\,.
\end{equation}
This motivates to introduce the momenta conjugate to the
$U_\mu(x)$ as $p_{\mu}^j(x)\in\mathbb{R}\,,\ j=1,2,3$. The elementary
update steps then read as follows
\begin{equation}
  \label{eq:elemupdate}
  \begin{split}
    p_{\mu}^j(x)(\tau+\Delta\tau) &= p_{\mu}^j(x)(\tau) + \Delta\tau
    D_j\s\,,\qquad j=1,2,3\,,\\
    U_\mu(x)(\tau+\Delta\tau) &= \exp\left[\ii\Delta\tau\,
    \sum_j p_{\mu}^j(x)(\tau+\Delta\tau/2)\sigma_j\right]\ U_\mu(x)(\tau)\,.\\
  \end{split}
\end{equation}
In order to study the response of the algorithm to increasing
reversibility violations, we deliberately round on the right hand
sides of \eqs{eq:elemupdate} to $d$ significant decimal digits. 
To be precise, we replace \eqs{eq:elemupdate} by
\begin{equation}
  \label{eq:elemupdated}
  \begin{split}
    p_{\mu}^j(x)(\tau+\Delta\tau) &= p_{\mu}^j(x)(\tau) + \Delta\tau
    [D_j\s]_d\,,\qquad j=1,2,3\,,\\
    U_\mu(x)(\tau+\Delta\tau) &= \exp\left[\ii\Delta\tau\,
    \sum_j p_{\mu}^j(x)(\tau+\Delta\tau/2)\sigma_j\right]\ \left[U_\mu(x)(\tau)\right]_d\,.\\
  \end{split}
\end{equation}
\eqs{eq:elemupdate} guarantee the $U$-fields to stay in
SU$(2)$. However, with finite precision arithmetics this is only true
up to rounding errors. Hence, we apply at the end of each MD evolution
a projection to SU$(2)$ $P_{\mathrm{SU}(2)}$. This is in particular
important when $d<16$. This projection step is applied before the
accept/reject step, thus, $P_{\mathrm{SU}(2)}$ will affect
reversibility and measure conservation at the same level as before. 
All runs with $d<16$ have been started from a well equilibrated
($\sim 5000$ trajectories) configuration of a run without rounding.

If not specified otherwise, the trajectory length is always chosen to
be $\tau=1$. This holds for all $\beta$-values and volumes. 
As random number generator we use the
Mersenne Twister algorithm~\cite{Matsumoto:1998:MTE:272991.272995}
implemented in the C++ standard library.
The SU$(2)$ simulation code is publicly available~\cite{urbach:su2} and
so is the analysis code~\cite{urbach:hadron}.

\subsection{Observables}

During the run of the HMC we will measure observables on each
trajectory. These are first of all the plaquette expectation value
reading
\begin{equation}
  \langle P\rangle\ =\ \frac{1}{6 L^3 T}\langle \sum_{x\in\Lambda}\sum_{\mu<\nu}
  \tr\, U_{\mu\nu}(x)\rangle\,. 
\end{equation}
The plaquette expectation value is one of the observables measurable
with very high statistical accuracy and hence a good candidate for
possible deviations. In addition to the plaquette itself, we also
measure its integrated autocorrelation time $\tau_\mathrm{int}(\langle
P\rangle)$ using the methods described in Ref.~\cite{Wolff:2003sm}. 

Next, we measure of course $\dH$ for each
trajectory, which gives access to $\langle \dH\rangle$ and $\langle
\exp(-\dH)\rangle$. The latter two are important to check whether
\eq{eq:edH} and \eq{eq:edH2} are fulfilled. It turns out that $\dH$
shows no autocorrelation, as one would expect. Another quantity we
measure for each trajectory is acceptance. From this we quantify the
acceptance rate $P_\mathrm{acc}$ in percent. 

More observables are measured only with a frequency of $100$
trajectories. First of all, we measure $\ddH$ by integrating backward
in time. It turns out that $\ddH$ is to a good approximation Gaussian
distributed with mean zero and standard deviation $\sd(\ddH)$, the
latter of which depends directly on the number of significant digits
used in the force calculation. Hence, $\sd(\ddH)$ will be used as a
measure for reversibility violations. 

The plaquette represents the smallest closed Wilson loop which can be
built on the lattice. As additional observables we consider planar
Wilson loops of extension $t\times r$
\begin{equation}
  C(t, r)\ =\ \frac{1}{3L^3T}\langle\sum_{x\in\Lambda}\sum_{\mu\neq 0} \tr\,
  U_{\mu}^{t, r} (x)\rangle\,.
\end{equation}
Here we denote the planar Wilson loop in spatial direction $\mu$ and
with time extent $t$ and spatial extent $r$ by $U_{\mu}^{t, r}$. 
$C(t, r)$ decays at fixed $r$ exponentially at large $t$ like
\begin{equation}
  C(t, r) \ \propto\ \exp(-V(r) t)
\end{equation}
with $V(r)$ the so-called static quark potential at spatial distance $r$.

\subsection{Lattice Scales}

\begin{table}[t]
  \centering
  \begin{tabular*}{.4\textwidth}{@{\extracolsep{\fill}}lrrr}
    \hline
    $\beta$ & $t_0/a^2$ & $s_0/a$ & $N_\mathrm{meas}$ \\
    \hline
    $2.3$ & $1.737(06)$ & $1.318(2)$ & $131$ \\
    $2.4$ & $2.790(23)$ & $1.670(7)$ & $112$\\
    $2.5$ & $5.038(36)$ & $2.245(8)$ & $130$\\
    \hline
  \end{tabular*}
  \caption{Gradient flow scales $t_0$ and $s_0=\sqrt{t_0}$ for the
    $\beta$-values in lattice units for the $\beta$-values used in
    this study. We also give the number of well separated
    configurations $N_\mathrm{meas}$ we measured the scales on.} 
  \label{tab:a}
\end{table}

SU$(2)$ gauge theory has been studied in the literature over many decades
using lattice techniques, starting with the famous paper by
Creutz~\cite{Creutz:1980zw} from 1980. Hence, scaling variables have
been determined, see for instance
Refs.~\cite{Perantonis:1988uz,Huntley:1985ts,Cardoso:2010di}. Still,
here we rely on the gradient flow~\cite{Luscher:2010iy}, recently
studied for SU$(2)$ Yang-Mills theory in Ref.~\cite{Berg:2016wfw}. 

We follow the notation and the
definitions of Ref.~\cite{Luscher:2010iy} and use the symmetric
definition of the energy density $E_\mathrm{sym}$. But since we work
in SU$(2)$, we use the following defining equation for the scale
$t_0$
\begin{equation}
  \label{eq:t0}
  t^2 \langle E_\mathrm{sym}(t)\rangle |_{t=t_0} = 0.1
\end{equation}
where $t$ is the so-called flow time. Note that we chose $0.1$ instead
of original SU$(3)$ value $0.3$ in \eq{eq:t0} following the reasoning in
Ref.~\cite{DeGrand:2017gbi}. In addition we define the length scale
$s_0$ via 
\begin{equation}
  \label{eq:s0}
  s_0 = \sqrt{t_0}\,.
\end{equation}
The choice of $\beta$-values used in this paper are motivated by the
requirement to be in the scaling region.
The values for $t_0/a^2$ and $s_0/a$ we have determined for these
$\beta$-values are compiled in \tab{tab:a}.
The precision of the scales is not central to the results of this
study, thus, we did not spend too many resources to obtain very
precise results. The configurations used for determining $t_0$ were
separated by at least $500$ HMC trajectories and, hence, free of
autocorrelation. For more details see the appendix.

The ratios of our $s_0$-values, given in \tab{tab:a}, can be
compared to the results presented in Ref.~\cite{Berg:2016wfw}. Only
roughly, because in Ref.~\cite{Berg:2016wfw} scales have been
determined for $\beta=2.3$, $\beta=2.43$ and $\beta=2.51$. Still, the
agreement is reasonable.

%% file: results.tex
\section{Results}

The statistical analysis of the Markov chains is performed using the
so-called $\Gamma$-method described in Ref.~\cite{Wolff:2003sm}. In
this way we include autocorrelation effects in the estimate of the
standard error by estimating the integrated autocorrelation time
$\tau_\mathrm{int}$ of the observable in question. This analysis is
double checked using a blocked bootstrap procedure, for which we find
consistent results.

\subsection{Results for $\beta=2.3$}

\begin{table}[t]
  \centering
  \begin{tabular*}{1.\textwidth}{@{\extracolsep{\fill}}llrrrrrrrr}
    \hline\hline
    Int & $d$ & $N_\mathrm{traj}$ & $\langle P\rangle$ & $\tau_\mathrm{int}(P)$ &
    $\langle\exp(-\Delta H)\rangle$ & $\langle\Delta H\rangle$ &
    $\sd(\delta\Delta H)$ & $\rho$ & $P_\mathrm{acc}$ \\
    \hline\hline
    \input{tableb23L16}  
    \hline\hline
  \end{tabular*}
  \caption{Results at $\beta=2.3$ and $L=16$, $T=32$. This can be
    compared to a Metropolis algorithm with $\langle P\rangle =
    0.602266(6)$.}
  \label{tab:resb23L16}
\end{table}

At the coarsest lattice spacing corresponding to $\beta=2.3$ we have
performed runs for three different spatial volumes $L=12, 16, 20$ and
a variety of significant digits $d$. We also compared the LF with the
OMF4 integration scheme. 

The runs and results for the observables $\langle P\rangle$, $\dH$
related and $\ddH$ are compiled for the different integration schemes
and different $d$-values in \tab{tab:resb23L16}, \tab{tab:resb23L12}
and \tab{tab:resb23L20}. For better readability we have moved most of
the tables to the appendix, apart from \tab{tab:resb23L16}. We quote
'$-$' for $d$ if we run in double precision and perform no rounding. It
roughly corresponds to $d=16$. 

For $L=16$ we have carried out a comparison to a Metropolis
algorithm, which yielded $\langle P\rangle = 0.602266(6)$ agreeing
perfectly within statistical errors with the double precision 
HMC run, either with LF or OMF4 integration scheme. 

The values of $d$ have been chosen as follows: we first determined the
value of $d$ where the HMC becomes instable. For $L=16$ and $L=12$ this
was the case for $d=3$, see \tab{tab:resb23L16} and
\tab{tab:resb23L12}, respectively. This instability manifests itself in a
significant increase in $\langle \dH\rangle$ compared to the run without
rounding, leading also to large drop in $P_\mathrm{acc}$. These runs
are clearly not reliable anymore, but also clearly identifiable as not
reliable. That the plaquette expectation value is still roughly in line
comes from the combination of low acceptance rate with an equilibrated
initial gauge configuration.

\begin{figure}[t]
  \centering
  \subfigure{\includegraphics[width=.48\linewidth,page=7]{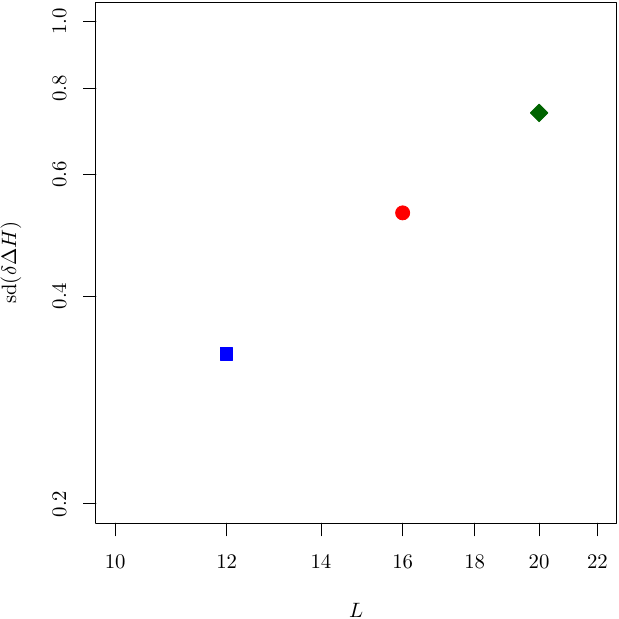}}
  \subfigure{\includegraphics[width=.48\linewidth,page=9]{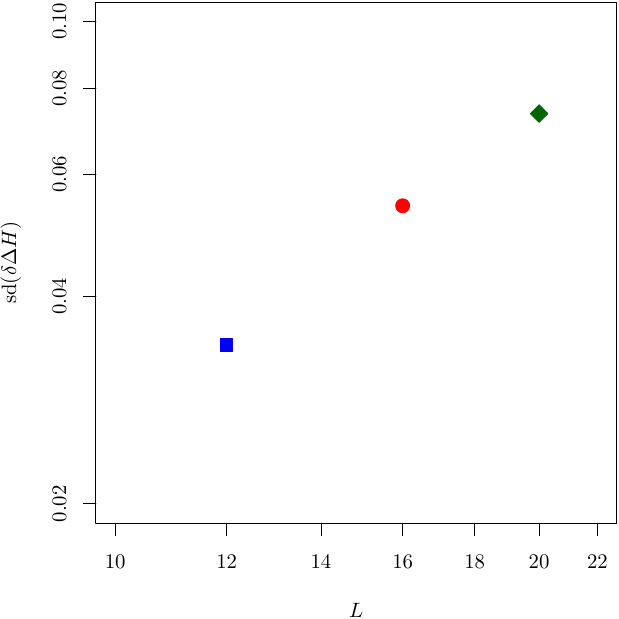}}
  \caption{$\sd(\ddH)$ as a function of $L$ in a double logarithmic
    plot for $\beta=2.3$ and $L=12,16,20$ with the OMF4 integration
    scheme. Left: $d=4$. Right: $d=5$. Note the factor $10$ difference
    in the scale of the $y$-axes.
  }
  \label{fig:depL}
\end{figure}

Looking at the $d$ dependence of $\sd(\ddH)$, we find to a good
approximation
\begin{equation}
  \log_{10}(\sd(\ddH))\ =\ c_d(L, \mathcal{I})\,d\,.
\end{equation}
Therefore, we will replace $d$ by $\sd(\ddH)$ as a measure of
reversibility violation. The coefficient $c_d$ depends on the volume
and the details of the integration scheme.
The dependence of $c_d$ on $L$ is shown in \fig{fig:depL}, in the left
panel for $d=4$ and in the right panel for $d=5$. $c_d$ turns out to
be proportional to $L^\gamma$, with $\gamma\sim 3/2$,
i.e. $c_d\propto\sqrt{L^3}$. This dependence is actually na{\"\i}vely
expected for $\sd(\ddH)$.

Let us now turn to the other observables quoted in
\tab{tab:resb23L12}, \tab{tab:resb23L16} and
\tab{tab:resb23L20}. First of all, in not one of the different runs
with $d<16$ a significant deviation of the plaquette expectation value
compared to the run without rounding could be detected.

However, in $\langle \exp(-\dH)\rangle$, $\langle\dH\rangle$ and
$P_\mathrm{acc}$ we observe deviations as $d$ is being decreased. 
Up to values $\sd(\ddH)\approx 0.1$, $\langle
\exp(-\dH)\rangle$ is compatible with one, as expected for the HMC. At
the same time $\langle \dH\rangle$ and $P_\mathrm{acc}$ are compatible
within errors with the results without rounding. For
$\sd(\ddH)\gtrsim 0.1$, we observe significant deviations in all
three observables. 
Starting with $\sd(\ddH)\approx 0.1$ we also observe
that the correlation  
\begin{equation}
  \rho\ =\ \operatorname{Cor}(\dH,\, \ddH) 
\end{equation}
starts to increase to values around $0.5$. This is an indication that
the actual value of $\dH$ is significantly influenced by the
reversibility violation, thus leading to an incorrectly sampled
probability distribution. 

\subsection{Dependence on the Lattice Spacing}

For studying the lattice spacing dependence, we study ensembles at
$\beta=2.3$, $\beta=2.4$ and $\beta=2.5$. Using the
length scale $s_0$, we keep the physical volume approximately fixed by
using $L/a=16$ at $\beta=2.3$, $L/a=20$ at $\beta=2.4$ and $L/a=24$ at
$\beta=2.5$. The results for $\beta=2.4$ with $L=20$ and $\beta=2.5$
with $L=24$ are
summarised in \tab{tab:resb24} and \tab{tab:resb25},
respectively. Results for an additional volume for $\beta=2.5$ with
$L=20$ are compiled in \tab{tab:resb25L20}.
The results for $\beta=2.3$ have been discussed previously and can be
found in \tab{tab:resb23L16}. 

\begin{figure}[t]
  \centering
  \subfigure{\includegraphics[width=.48\linewidth,page=2]{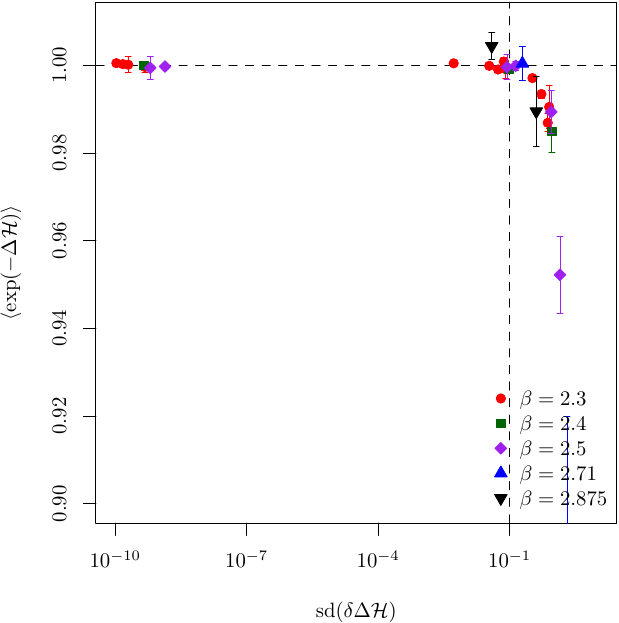}}
  \subfigure{\includegraphics[width=.48\linewidth,page=3]{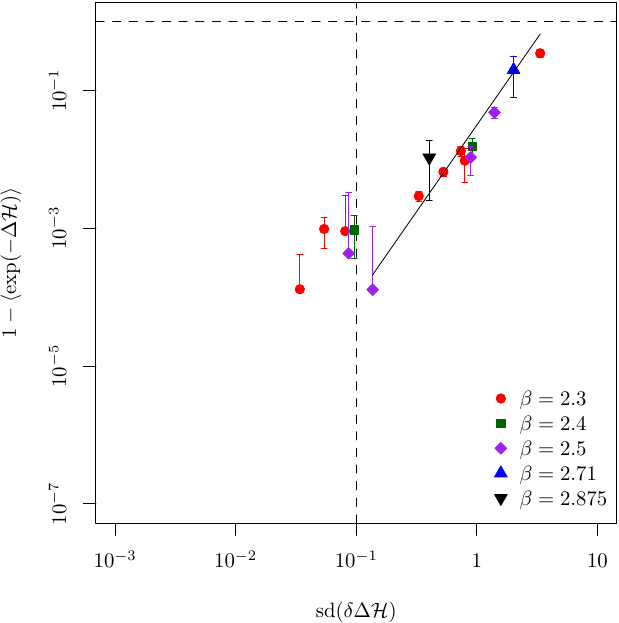}}
  \caption{Left:
    $\langle\exp(-\dH)\rangle$ as a function of $\sd(\ddH)$. Right:
    $1-\langle\exp(-\dH)\rangle$ as a function of $\sd(\ddH)$ in a
    double logarithmic plot. Data for all $\beta$-values, volumes and
    integration schemes are shown together. In the right panel also a
    fit to the data is shown in the range indicated by the line.}
  \label{fig:hmcalledH}
\end{figure}

First we discuss the results for $\langle \exp(-\dH)\rangle$ as a
function of $\sd(\ddH)$ by including all available 
$\beta$-values and volumes. This is shown in \fig{fig:hmcalledH} where in
the left panel $\langle \exp(-\dH)\rangle$ is plotted as a function of
$\sd(\ddH)$ with logarithmic $x$-axis and in the right panel
$1-\langle \exp(-\dH)\rangle$ as a function of $\sd(\ddH)$ with both
axes logarithmic. We find that all
the points fall on a universal curve within error bars. In the double
logarithmic plot (right panel) it is visible that the dependence of
$1-\langle \exp(-\dH)\rangle$ on $\sd(\ddH)$ is like
\begin{equation}
  1-\langle\exp(-\dH)\rangle\ \propto\ \sd(\ddH)^\delta + c_3
\end{equation}
with some exponent $\delta$ and a constant shift $c_3$. A fit to the
data points with $\sd(\ddH)>0.1$ reveals 
\[
\delta\ =\ 2.6(3)\,
\]
and a value for $c_3$ significantly non-zero. The origin of the actual
value of $\delta$ and in particular the non-zero shift $c_3$ is not
clear as one would expect $c_3$ to be zero if reversibility was
restored smoothly. At this point it is just an empirical finding.

\begin{figure}[t]
  \centering
  \subfigure{\includegraphics[width=.48\linewidth,page=4]{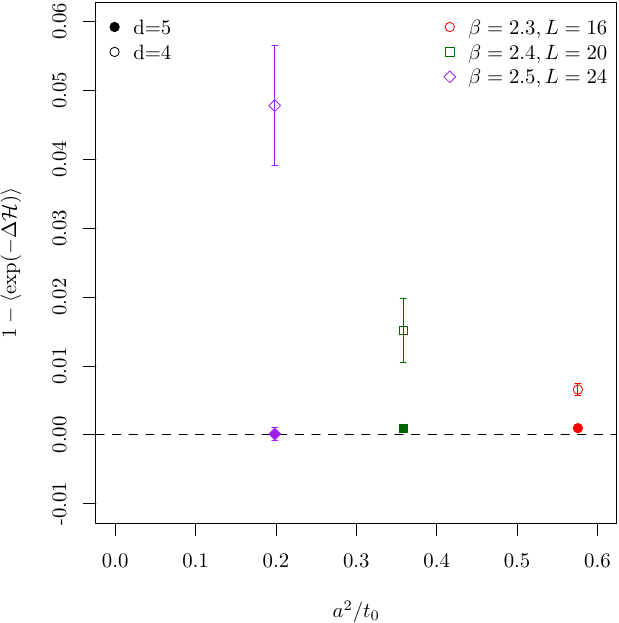}}
  \subfigure{\includegraphics[width=.48\linewidth,page=5]{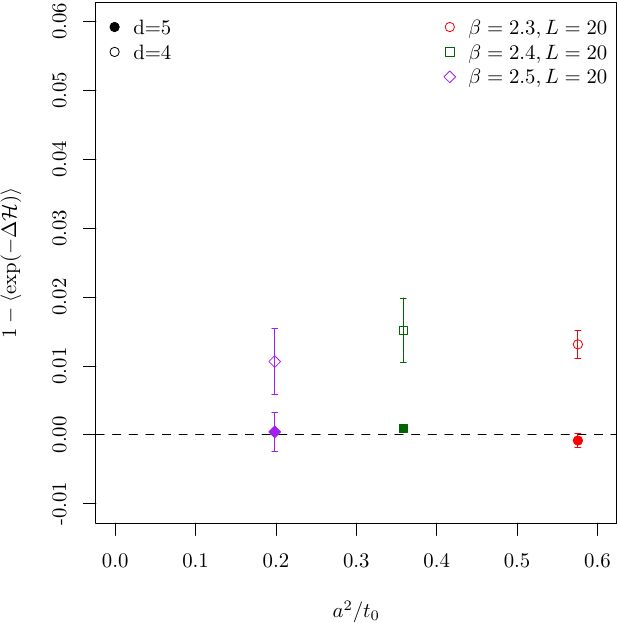}}
  \caption{$1-\langle\exp(-\dH)\rangle$ as a function of $a^2/t_0$. 
    Left: physical volume fixed. Right: $L/a$ fixed.}
  \label{fig:t0dep}
\end{figure}

In \fig{fig:t0dep} we show $1-\langle\exp(-\dH)\rangle$ as a function
of the gradient flow scale $a^2/t_0$ both for $d=4$ and $d=5$. In the
left panel we keep the physical volume, i.e. $L/s_0$ approximately
fixed. In the right panel we keep the number of lattice points $L/a$
fixed. In the latter case we observe no dependence on $a^2/t_0$,
neither for $d=5$ nor for $d=4$. For the case of fixed $L/s_0$ and
$T/s_0$ we observe an increase for $d=4$ towards smaller $a^2/t_0$
values.  

\subsection{Wilson Loops at $\beta=2.5$}

\begin{figure}[t]
  \centering
  \subfigure{\includegraphics[width=.48\linewidth,page=1]{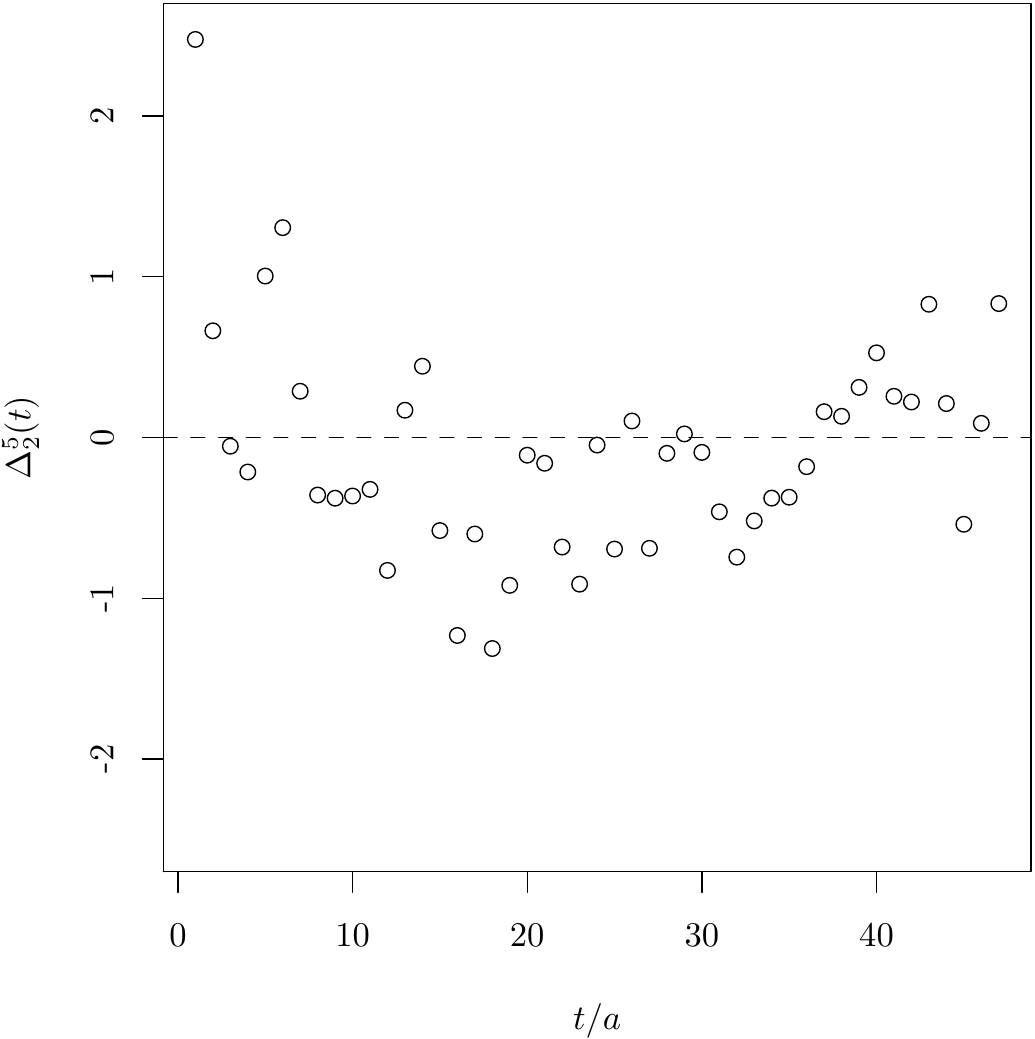}}
  \subfigure{\includegraphics[width=.48\linewidth,page=2]{Wloop}}\\
  \subfigure{\includegraphics[width=.48\linewidth,page=3]{Wloop}}
  \subfigure{\includegraphics[width=.48\linewidth,page=4]{Wloop}}

  \caption{$\Delta_r^d$ as a function
    of Euclidean time $t/a$ measured at $\beta=2.5$ with $L=24$ and
    $T=48$. Left: $d=5$. Right: $d=4$. Upper row: $r/a=2$. Lower row: $r/a=8$}
  \label{fig:Wloop}
\end{figure}

We have studied planar Wilson loops of fixed spatial extend $r$ as a
function of $t$ for $\beta=2.5$. We have computed the loops on
configurations generated without deliberate rounding, with $d=5$ and
$d=4$, see \tab{tab:resb25}. Next we define the following normalised
differences
\begin{equation}
  \Delta^r_d(t)\ =\ \frac{C^d(t,r) -
    C^0(t,r)}{\sqrt{\mathrm{d}C^d(t,r)^2 + \mathrm{d}C^0(t,r)^2}}\,. 
\end{equation}
Here we denote the standard error of $C^d(t,r)$ with $\mathrm{d}C^d(t,r)$. 
In \fig{fig:Wloop} we plot $\Delta_r^d(t)$ as a function of $t/a$. 
In the upper row we plot data for $r/a=2$, in the lower one for
$r/a=8$. The left column corresponds to $d=5$, the right one to $d=4$. 
Note that for $r/a=2$ the signal is lost in the noise at around
$t/a=20$ and for $r/a=8$ around $t/a=10$.

For $r/a=2$, we observe a number of values of $\Delta_2^d(t)$ with
modulus around $2$. Still, there is no single $t$-value where the
deviation from $0$ is significant. In agreement with the results for
the plaquette expectation value we, hence, find also for the exemplary
Wilson loops we looked at no sign of a deviation due to reversibility
violations.

%% file: tableb23L16.tex
LF & $-$ & $130000$ & $0.60225(1)$ & $6.3(2)$ & $1.0001(18)$ & $0.2008(17)$ & $2.0\cdot10^{-10}$ & $-0.02$ & $75$ \\ 
LF & $5$ & $100737$ & $0.60226(1)$ & $5.9(2)$ & $0.9991(21)$ & $0.2048(20)$ & $0.08087973$ & $0.09$ & $75$ \\ 
LF & $4$ & $98285$ & $0.60226(1)$ & $8.2(4)$ & $0.9905(50)$ & $0.6481(38)$ & $0.7943932$ & $0.42$ & $57$ \\ 
OMF4 & $-$ & $530000$ & $0.602264(3)$ & $5.8(1)$ & $1.00026(45)$ & $0.05312(44)$ & $1.5\cdot10^{-10}$ & $0$ & $87$ \\ 
OMF4 & $6$ & $420000$ & $0.602266(3)$ & $5.8(1)$ & $1.00044(51)$ & $0.05295(49)$ & $0.005266247$ & $0.01$ & $87$ \\ 
OMF4 & $5$ & $520000$ & $0.602266(3)$ & $5.8(1)$ & $0.99903(46)$ & $0.05541(45)$ & $0.05406281$ & $0.08$ & $87$ \\ 
OMF4 & $4$ & $480000$ & $0.602266(3)$ & $6.6(1)$ & $0.99343(89)$ & $0.16643(83)$ & $0.528$ & $0.46$ & $77$ \\ 
OMF4 & $3$ & $40000$ & $0.6023(1)$ & $251(71)$ & $0.174(45)$ & $11.337(30)$ & $5.238426$ & $0.59$ & $2$ \\ 

%% file: summary.tex
\section{Discussion and Summary}

The results presented in the last section indicate that -- at least
for SU$(2)$ gauge theory -- reversibility violations do not lead to
deviations in the physical observables studied here. This is
surprising, because for the observable $\exp(-\dH)$ with the
analytically known expectation value we observe such deviations. 

It turns out that good quantities to monitor reversibility are $\langle
\exp(-\dH)\rangle$ and $\sd(\ddH)$. One observes that $\sd(\ddH)$ is
directly proportional to the rounding errors introduced deliberately
in the HMC MD evolution. In the range of $\beta$-values studied here,
$\langle \exp(-\dH)\rangle$ turns out to be a universal function of
$\sd(\ddH)$, independent of integration scheme and problem size. With
$\sd(\ddH)\lesssim0.1$ no significant deviations of $\langle
\exp(-\dH)\rangle$ from one could be detected. For $\sd(\ddH)\gtrsim
0.1$ these deviations become significant and follow a power law in
$\sd(\ddH)$. It is very 
likely that with even larger statistical accuracy also for
$\sd(\ddH)<0.1$ significant deviations from one will be detectable.
However, they will be tiny.

Another important observation is the fact that reversibility violations
always lead to an increase in $\langle \dH\rangle$ towards positive
values. As a consequence, with too large violations the acceptance
rate drops significantly. That the reversibility violations are
largely responsible for the large $\dH$-values is indicated by the
fact that for $\sd(\ddH)\gtrsim 0.1$ the correlation between $\dH$ and $\ddH$
becomes significant. This indicates are large influence of the
reversibility violations on the accept/reject decision.

When changing $\beta$, deviations in $\langle \exp(-\dH)\rangle$ do
not depend on $\beta$ if the number of lattice points is kept
constant. In turn, when the physical volume is kept constant,
deviations increase towards the continuum limit. This could on the one
hand be an indication that the underlying Lyapunov exponent is not
varying much with $\beta$. Another possible reason could be that with
trajectory lengths of $\tau=1$ the system is 
still in the ``random walk'' regime and not yet in the regime where
deviations increase exponentially. The latter interpretation is
supported by the results of Ref.~\cite{Jansen:1995gz,Liu:1997fs}.

In summary, simulations with HMC should be safe as long as
$\sd(\ddH)<0.1$ and correlations between $\dH$ and $\ddH$ are
negligible. Those quantities are easy to monitor. In fact, $\dH$ is
available anyhow, because it is needed for the accept/reject
test. $\sd(\ddH)$ can be measured by performing reversibility tests
on, say, $\mathcal{O}(100)$ trajectories.
It remains to be seen whether the results found here for SU$(2)$ gauge
theory generalise to QCD with SU$(3)$ gauge symmetry and dynamical
fermions.

%% file: appendix.tex
\begin{appendix}

\section{Gradient Flow Scales}

\begin{figure}[t]
  \centering
  \subfigure{\includegraphics[width=.48\linewidth,page=1]{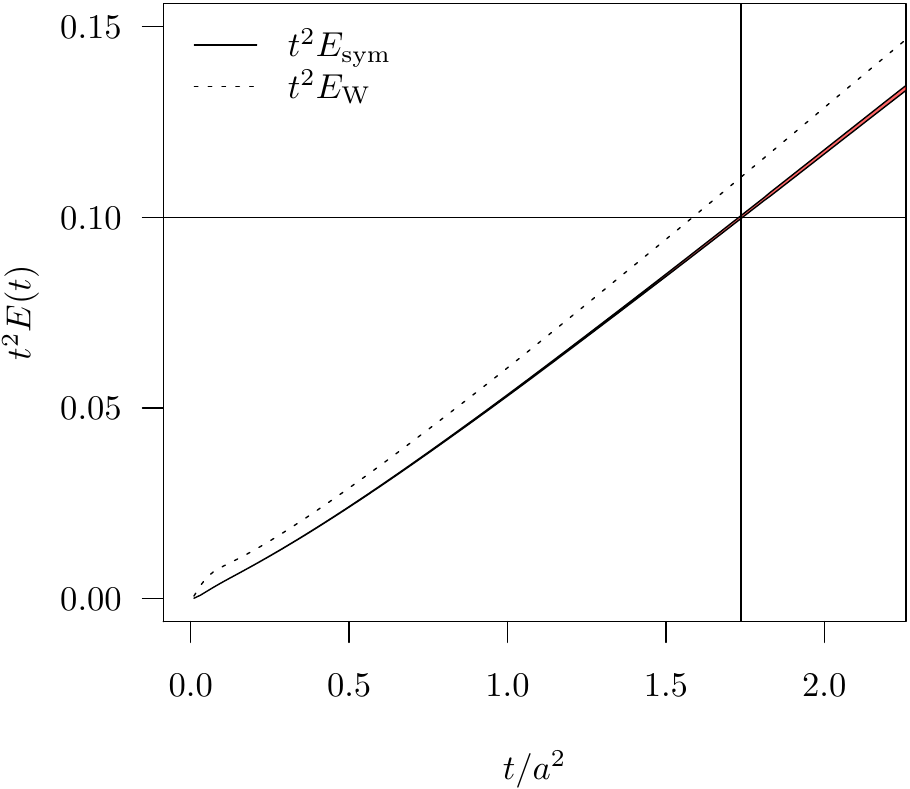}}
  \subfigure{\includegraphics[width=.48\linewidth,page=1]{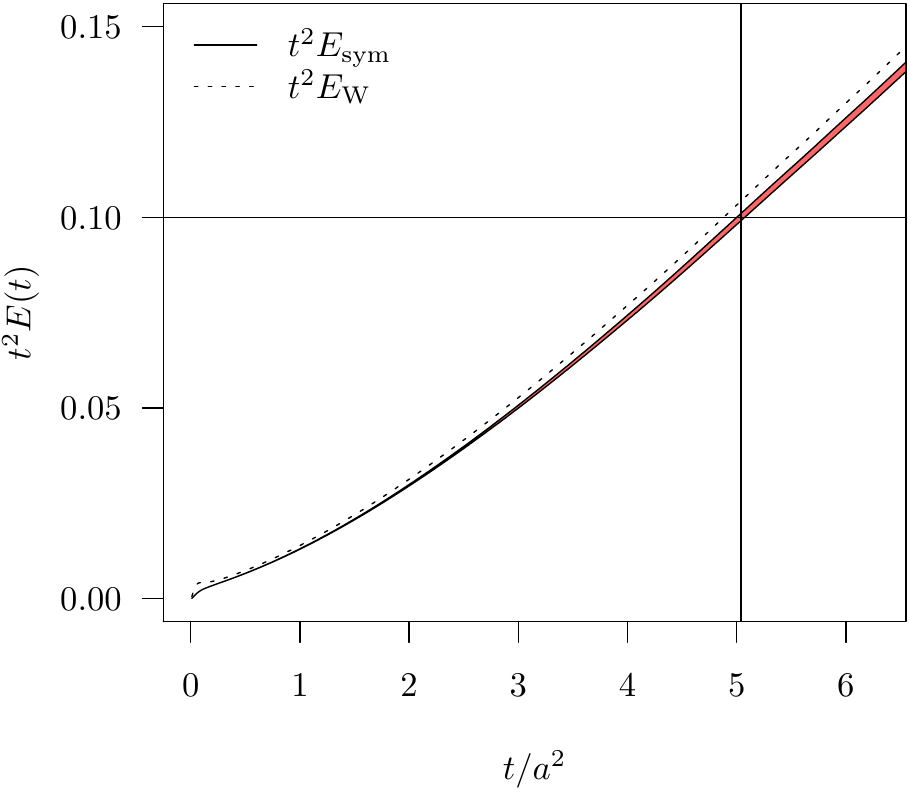}}
  \caption{Gradient flow for  $\beta=2.3$ (left) and $\beta=2.5$ (right).
  }
  \label{fig:gradflow}
\end{figure}

For determining the scales $t_0$ \eq{eq:t0} and $s_0$ \eq{eq:s0} we follow
the approach presented in the original paper by
L{\"u}scher~\cite{Luscher:2010iy}. The energy density $E$ can be
defined symmetrically as the sum over the four plaquettes attached to
a point $x$ (the clover definition). This one we will denote with
$E_\mathrm{sym}$. A second possibility is to use the action
\eq{eq:actionsu2}, which we denote as $E_\mathrm{W}$. For the exact factors see
Ref.~\cite{Luscher:2010iy}. We use $E_\mathrm{sym}$ to determine the
scales $t_0$ and $s_0$, because in Ref.~\cite{Luscher:2010iy} it was
found to have less lattice artefacts, and use $E_\mathrm{W}$ as a cross-check. 

In \fig{fig:gradflow} we show $t^2 E(t)$ as a function of the flow
time $t/a^2$ for $\beta=2.3$ (left panel) and $\beta=2.5$ (right
panel). The solid lines with error band correspond to $E_\mathrm{sym}$
and the dashed line to $E_\mathrm{W}$. The cross indicates the determination
of $t_0$ where $t^2 E(t)=0.1$. We observe differences between the two
definitions of $E$ which, however, decrease towards the continuum
limit, as expected. 

We remark here that we cannot quantitatively reproduce the results of
Ref.~\cite{Berg:2016wfw} for $\beta=2.3$. Our definition of
$E_\mathrm{sym}$ differs by a factor of two to the one from
Ref.~\cite{Berg:2016wfw}, but this factor is not sufficient to obtain
agreement. We remark that we have two independent implementations,
which agree. Moreover, we have a strong test of the derivative,
because it is used in the HMC as well. Apart from that the ratios of
scales agree with the ones from Ref.~\cite{Berg:2016wfw}, as far as
this can be judged due to not exactly identical $\beta$-values.

\FloatBarrier

\section{Data Tables}

\begin{table}[htbp]
  \centering
  \begin{tabular*}{1.\textwidth}{@{\extracolsep{\fill}}llrrrrrrrr}
    \hline\hline
    Int & $d$ & $N_\mathrm{traj}$ & $\langle P\rangle$ & $\tau_\mathrm{int}(P)$ &
    $\langle\exp(-\Delta H)\rangle$ & $\langle\Delta H\rangle$ &
    $\sd(\delta\Delta H)$ & $\rho$ & $P_\mathrm{acc}$ \\
    \hline\hline
    \input{tableb23L12}  
    \hline\hline
  \end{tabular*}
  \caption{Results at $\beta=2.3$ and $L=12$, $T=32$.}
  \label{tab:resb23L12}
\end{table}

\begin{table}[htbp]
  \centering
  \begin{tabular*}{1.\textwidth}{@{\extracolsep{\fill}}llrrrrrrrr}
    \hline\hline
    Int & $d$ & $N_\mathrm{traj}$ & $\langle P\rangle$ & $\tau_\mathrm{int}(P)$ &
    $\langle\exp(-\Delta H)\rangle$ & $\langle\Delta H\rangle$ &
    $\sd(\delta\Delta H)$ & $\rho$ & $P_\mathrm{acc}$ \\
    \hline\hline
    \input{tableb23L20}  
    \hline\hline
  \end{tabular*}
  \caption{Results at $\beta=2.3$ and $L=20$, $T=32$.}
  \label{tab:resb23L20}
\end{table}

\begin{table}[htbp]
  \centering
  \begin{tabular*}{1.\textwidth}{@{\extracolsep{\fill}}llrrrrrrrr}
    \hline\hline
    Int & $d$ & $N_\mathrm{traj}$ & $\langle P\rangle$ & $\tau_\mathrm{int}(P)$ &
    $\langle\exp(-\Delta H)\rangle$ & $\langle\Delta H\rangle$ &
    $\sd(\delta\Delta H)$ & $\rho$ & $P_\mathrm{acc}$ \\
    \hline\hline
    \input{tableb24L20}  
    \hline\hline
  \end{tabular*}
  \caption{Results at $\beta=2.4$ and $L=20$, $T=40$.}
  \label{tab:resb24}
\end{table}

\begin{table}[htbp]
  \centering
  \begin{tabular*}{1.\textwidth}{@{\extracolsep{\fill}}llrrrrrrrr}
    \hline\hline
    Int & $d$ & $N_\mathrm{traj}$ & $\langle P\rangle$ & $\tau_\mathrm{int}(P)$ &
    $\langle\exp(-\Delta H)\rangle$ & $\langle\Delta H\rangle$ &
    $\sd(\delta\Delta H)$ & $\rho$ & $P_\mathrm{acc}$ \\
    \hline\hline
    \input{tableb25L20}  
    \hline\hline
  \end{tabular*}
  \caption{Results at $\beta=2.5$ and $L=20$, $T=40$.}
  \label{tab:resb25L20}
\end{table}

\begin{table}[htbp]
  \centering
  \begin{tabular*}{1.\textwidth}{@{\extracolsep{\fill}}llrrrrrrrr}
    \hline\hline
    Int & $d$ & $N_\mathrm{traj}$ & $\langle P\rangle$ & $\tau_\mathrm{int}(P)$ &
    $\langle\exp(-\Delta H)\rangle$ & $\langle\Delta H\rangle$ &
    $\sd(\delta\Delta H)$ & $\rho$ & $P_\mathrm{acc}$ \\
    \hline\hline
    \input{tableb25L24}  
    \hline\hline
  \end{tabular*}
  \caption{Results at $\beta=2.5$ and $L=24$, $T=48$.}
  \label{tab:resb25}
\end{table}

\end{appendix}

%% file: tableb23L12.tex
OMF4 & $-$ & $300009$ & $0.60226(1)$ & $5.4(1)$ & $1.00048(38)$ &
$0.02191(38)$ & $1.0\cdot 10^{-10}$ & $0.03$ & $92$ \\ 
OMF4 & $5$ & $555300$ & $0.602264(4)$ & $5.4(1)$ & $0.99987(28)$ & $0.02304(28)$ & $0.03400932$ & $0.08$ & $91$ \\ 
OMF4 & $4$ & $591001$ & $0.602254(4)$ & $5.7(1)$ & $0.99708(49)$ & $0.07037(48)$ & $0.3297102$ & $0.46$ & $85$ \\ 
OMF4 & $3$ & $110000$ & $0.60227(3)$ & $48(5)$ & $0.655(45)$ & $4.806(11)$ & $3.351697$ & $0.57$ & $12$ \\ 

%% file: tableb23L20.tex
OMF4 & $-$ & $170800$ & $0.602258(4)$ & $6.1(2)$ & $0.9996(11)$ & $0.1039(11)$ & $4.8\cdot10^{-10}$ & $-0.02$ & $82$ \\ 
OMF4 & $5$ & $216501$ & $0.602261(3)$ & $6.2(2)$ & $1.0008(10)$ & $0.1053(10)$ & $0.07353326$ & $0.08$ & $82$ \\ 
OMF4 & $4$ & $206501$ & $0.602254(4)$ & $7.3(2)$ & $0.9869(20)$ & $0.3276(18)$ & $0.7368223$ & $0.48$ & $69$ \\ 

%% file: tableb24L20.tex
OMF4 & $-$ & $67401$ & $0.630000(4)$ & $4.7(2)$ & $0.99996(37)$ & $0.00551(37)$ & $4.5\cdot10^{-10}$ & $0.08$ & $96$ \\ 
OMF4 & $5$ & $50000$ & $0.629993(5)$ & $4.7(2)$ & $0.99906(58)$ & $0.01005(58)$ & $0.0958693$ & $0.38$ & $94$ \\ 
OMF4 & $4$ & $50000$ & $0.63000(1)$ & $7.6(5)$ & $0.9848(47)$ & $0.3818(39)$ & $0.91566$ & $0.53$ & $66$ \\ 

%% file: tableb25L20.tex
OMF4 & $-$ & $73900$ & $0.651965(4)$ & $4.9(2)$ & $0.9994(27)$ & $0.2197(24)$ & $6.3\cdot10^{-10}$ & $-0.02$ & $74$ \\ 
OMF4 & $5$ & $69233$ & $0.651966(4)$ & $4.3(2)$ & $0.9996(28)$ & $0.2232(25)$ & $0.08636778$ & $0.09$ & $74$ \\ 
OMF4 & $4$ & $80000$ & $0.651962(4)$ & $6.0(3)$ & $0.9894(48)$ & $0.5349(38)$ & $0.8890045$ & $0.48$ & $61$ \\ 

%% file: tableb25L24.tex
OMF4 & $-$ & $74901$ & $0.651967(2)$ & $3.7(1)$ & $0.99973(57)$ & $0.01409(56)$ & $1.4\cdot10^{-09}$ & $-0.05$ & $93$ \\ 
OMF4 & $5$ & $46500$ & $0.651965(3)$ & $3.8(2)$ & $0.99987(94)$ & $0.02163(93)$ & $0.136749$ & $0.32$ & $92$ \\ 
OMF4 & $4$ & $45000$ & $0.651970(4)$ & $6.7(4)$ & $0.9522(87)$ & $0.8515(63)$ & $1.398589$ & $0.55$ & $51$ \\ 